\documentclass{article}

\usepackage{euscript}

\usepackage{latexsym}
\usepackage{amsmath, amssymb,graphics}
\usepackage[all]{xy}

\usepackage{fullpage}

\newtheorem{theorem}{Theorem}
\newtheorem{definition}[theorem]{Definition}

\newtheorem{proposition}[theorem]{Proposition}

\newtheorem{observation}[theorem]{Observation}

\renewcommand{\(}{\begin{equation*}}
\renewcommand{\)}{\end{equation*}}
\newcommand{\bea}{\begin{eqnarray*}}
\newcommand{\eea}{\end{eqnarray*}}
\newcommand{\R}{{\mathbb R}}
\newcommand{\C}{{\mathbb C}}
\newcommand{\Z}{{\mathbb Z}}
\newcommand{\Q}{{\mathbb Q}}

\newcommand{\cL}{\ensuremath{\mathcal L}}

\newcommand{\bo}{\raise-1mm\hbox{\Large$\Box$}}              
\def\O{\ensuremath{{\cal O}}}

\newcommand{\cS}{{\mathcal S}}

\def\O{\ensuremath{{\cal O}}}

\newcommand{\beq}{\begin{equation}}
\newcommand{\eeq}{\end{equation}}

\numberwithin{equation}{section}

\renewcommand{\(}{\begin{equation}}
\renewcommand{\)}{\end{equation}}

\newcommand{\CC}{{\mathbb C}}

\newcommand{\CP}{\CC \text{P}}

\def\R{{\mathbb R}}
\def\Z{{\mathbb Z}}
\def\Q{{\mathbb Q}}
\def\C{{\mathbb C}}

\def\1{{\bf 1}}

\def\<{\langle}
\def\>{\rangle}

\def\O{{\cal O}}

\numberwithin{equation}{section}

\renewcommand{\(}{\begin{equation}}
\renewcommand{\)}{\end{equation}}

\begin{document}

\begin{titlepage}


\vspace{2em}
\def\thefootnote{\fnsymbol{footnote}}

\begin{center}
{\Large\bf 
M-theory, the signature theorem, and geometric invariants
}
\end{center}
\vspace{1em}

\begin{center}
\large Hisham Sati 
\footnote{e-mail: {\tt
hsati@math.umd.edu}}
\end{center}

\begin{center}
Department of Mathematics\\
University of Maryland\\
College Park, MD 20742 
\end{center}

\vspace{0em}
\begin{abstract}
\noindent
The equations of motion and the Bianchi identity of the C-field in M-theory 
are encoded in terms of the signature operator. We then 
 reformulate the topological part of the action in M-theory using the signature,
 which leads to connections to the geometry of the underlying manifold, 
 including positive scalar curvature. 
This results in  
a variation on the
miraculous cancellation formula of Alvarez-Gaum\'e and Witten in twelve dimensions
and leads naturally to the Kreck-Stolz $s$-invariant in eleven dimensions. 
Hence M-theory detects 
diffeomorphism type of eleven-dimensional (and seven-dimensional) manifolds,
and in the restriction to parallelizable manifolds classifies topological eleven-spheres.  
Furthermore, requiring the phase of the partition function to be anomaly-free
imposes restrictions on allowed values of the $s$-invariant.
Relating to string theory in ten dimensions amounts to 
viewing the bounding theory as a disk bundle,
for which we study 
the corresponding phase in this formulation.



\end{abstract}

\end{titlepage}

\tableofcontents

\section{Introduction}

In this paper we show that M-theory encodes geometric invariants
via the signature. 
The index theorem is a powerful tool in characterizing anomalies of physical 
theories. The simplest example of an index theorem in even dimensions 
is the index of the de Rham  operator resulting in 
the Gauss-Bonnet theorem, which gives the Euler characteristic as a topological 
invariant. Due to the presence of spinors in a supersymmetric theory, 
it is very common to use the Dirac operator, whose index leads to the 
$\widehat{A}$-genus as a topological invariant.  In odd dimensions, there
is a geometric/analytical correction term, namely the eta-invariant 
in the Atiyah-Patodi-Singer (APS) index theorem
\cite{APS}. 
In this paper we consider instead, in the context of M-theory, 
 the 
 signature
 operator on differential forms, 
whose index is the signature \cite{Hir}. 
In the presence of an odd-dimensional boundary, 
the APS index theorem for the signature 
expresses the signature of a Riemannian manifold with
boundary in terms of the integral of the Hirzebruch L-polynomial 
and the eta-invariant of the boundary \cite{APS1}.

\vspace{3mm}
The C-field in eleven-dimensional 
M-theory with field strength $G_4$ has a dual field 
$G_7$, which is just the Hodge dual at the level of differential forms.  
Due to the structure of the equation of motion of the C-field, 
$G_7$ can also be viewed as a potential with a field strength 
$G_8$ \cite{DFM} \cite{S1} \cite{S2} \cite{S3} \cite{S6}.
We show how the signature operator in eleven dimensions
encodes the dynamics of the C-field and its dual. We also consider the 
index of this operator.
The harmonic part of the C-field $C_3$ is already studied
 in \cite{tcu}. In a complementary 
way, we consider here the
harmonic part of the field strength $G_4$. 
This is done in section \ref{sec 12 11}, where we provide 
observations which serve as preparation for the main discussion 
in section \ref{sec inv}.

\vspace{3mm}
Witten \cite{Flux} wrote the topological part of the action in M-theory on
a Spin eleven-manifold $Y^{11}$,
namely the combination of the Chern-Simons term and the one-loop 
term, using index theory.
This is done by lifting to the `bounding theory' on a Spin twelve-manifold 
$Z^{12}$ and involves an index of the Dirac operator coupled to an $E_8$ bundle 
as well as the index of the Rarita-Schwinger operator, that is the 
Dirac operator coupled to the virtual vector bundle $TZ^{12} - 4 \O$.
The subtraction of four copies of the trivial line bundle $-4 \O$ from the 
tangent bundle  
comes from the consideration of ghosts in eleven
dimensions. In section \ref{sec q} we give 
an alternative description of the topological part of the action, 
using the  
Hirzebruch signature 
theorem, and hence the Hirzebruch L-polynomial \cite{Hir}.  
This leads to 
a variant of the miraculous cancellation formula of 
 Alvarez-Gaum\'e
and Witten \cite{AW}
which we might call 
``quantum"
in the sense that ghosts coming from the path integral -- a quantum 
effect-- are accounted for.

\vspace{3mm}
The phase of the partition function in eleven dimensions (as opposed to 
twelve) involves the eta-invariants of the $E_8$ Dirac operator 
and of the Rarita-Schwinger operator. 
In section \ref{sec dif} we show that the
 above-mentioned reformulation in terms of the 
signature and the L-polynomial, when the $E_8$ bundle is trivial,
 leads essentially to the $s$-invariant of 
Kreck and Stolz \cite{KS}, defined in the rational numbers 
for positive scalar curvature metrics on 
our eleven-manifolds. Furthermore, absence of anomalies from the 
phase imposes a condition on the allowed values of the $s$-invariant. 
Issues of positive scalar curvature in M-theory in 
relation to the partition function are studied extensively in \cite{DMW-Spinc}.
The $s$-invariant requires the vanishing of the rational Pontrjagin classes $p_i$.
For $p_1$ this is weaker than requiring a String structure, the obstruction to 
which is $\frac{1}{2}p_1 \in H^4\Z$, because of possible 2-torsion. Similarly for 
$p_2$ this is weaker than requiring a Fivebrane structure \cite{SSS1} \cite{SSS2}, 
the obstruction to which is $\frac{1}{6}p_2 \in H^8\Z$, because of possible
2- and 3-torsion. 

\vspace{3mm}
The restriction of the $s$-invariant to parallelizable manifolds is given by the 
Eells-Kuiper invariant \cite{EK}. Since this invariant classifies topological spheres, we 
get in section \ref{sec dif} that M-theory classifies topological eleven-spheres. 
On the other hand, the extensions to the case when the $E_8$ bundle is 
no longer trivial suggests a possible generalization of the Kreck-Stolz invariant and which is 
defined for manifolds of positive scalar curvature together with a degree four 
cohomology class. To make our statements about M-theory will will 
rely on the corresponding constructions in \cite{KS}.

\vspace{3mm}
We also relate the geometric/analytical invariants in eleven dimensions 
 to type IIA string theory 
via dimensional reduction on the circle $S^1$ in section \ref{Sec s1}.  
We consider the 
adiabatic limit of the eta-invariant of the signature operator, 
as opposed to  that of the (twisted) Dirac operator considered previously 
in  \cite{MS} \cite{S-gerbe} \cite{DMW-Spinc}, and building on \cite{DMW}. 
The bounding theory is then taken on a disk bundle $Z^{12}$ 
over the ten-dimensional manifold of type IIA string theory.
The proof in \cite{APS1} of the index theorem with boundary 
assumes that the Riemannian manifold has a product metric near the boundary. 
For general manifolds, there is a correction form \cite{G},  
which should be used for disk bundles.
The signature of a disk bundle is given in terms of the integral 
of a characteristic class on the base manifold and a limiting eta-invariant
\cite{Ti}. We discuss this in section \ref{sec dis}.

\section{The signature (operator) in twelve and eleven dimensions}
\label{sec 12 11}

\subsection{The signature (operator) in twelve dimensions}
\label{sig 12}

\paragraph{The signature operator on closed twelve-manifolds.} 
Let $Z^{12}$ be an oriented Riemannian twelve-manifold. The de Rham operator $d$ and its
adjoint $d^*$ act on differential forms. The operator $d + d^*$ acts on the 
space $\Omega^*_Z$ of all differential forms and anticommutes with the involution
$\tau$ defined by $\tau \phi=-i^{p(p-1)}*_{12}\phi$ for $\phi \in \Omega^p_Z$
a $p$-form on $Z^{12}$. 
Denoting by $\Omega^{+}_Z$ and $\Omega^{-}_Z$ the $\pm$-eigenspaces of 
$\tau$, we have that $d+ d^*$ interchanges $\Omega^{+}_Z$ and $\Omega^{-}_Z$ and hence 
defines by restriction the signature operator $\sigma: \Omega^{+}_Z
\to \Omega^{-}_Z$. 

\vspace{3mm}
For $Z^{12}$ closed, Hodge theory gives the equality \cite{Hir}
\(
 {\rm sign}(Z^{12})={\rm index}(\sigma)=\int_{Z^{12}} L\;,
\)
where $L$ is the Hirzebruch L-polynomial and the right-hand side 
is 
the signature of the quadratic form on $H^6(Z^{12};\R)$ given by the
cup product. 
There is a bilinear form
on $H^{6} (Z^{12}) \otimes H^{6}(Z^{12}) \buildrel{\cup}\over{\longrightarrow}  \R$,
where the cup product $\cup$ is symmetric and nondegenerate, and the signature of $Z^{12}$ is 
$\sigma (Z^{12})= \sigma (\cup)$ with the following relevant properties

\begin{enumerate}
\item {\it Product}: $\sigma (M^{4} \times N^{8})= \sigma(M^{4}) \sigma (N^{8})$.
This will be useful in compactifications to four dimensions and to relating the 
corresponding secondary invariants in eleven dimensions to those in seven dimensions. 
\item {\it Bordism invariance}: If $Z^{12}= \partial W^{13}$ then $\sigma (Z^{12})=0$.
In this case 
the integral of the L-genus is zero. 

\end{enumerate}

\paragraph{Example: Calabi-Yau compactification.}
The signature can be used to derive consistency conditions on 
realistic compactifications \cite{Y}.
Consider the bounding theory on $Z^{12}$ taken to be a product 
$M^4 \times N^8$, where $M^4$ is a four-dimensional manifold and 
$N^8$ is a Calabi-Yau four-fold $N^8=X_{\C}^4$.
If $M^4$ is flat then the signature $\sigma (N^4)=0$ so that,
by the product property above, the signature of $Z^{12}$ is zero. 
If $M^4$ is not flat then $\sigma (M^4)$ can be  nonzero and 
$N^8$ can be taken so that
$\sigma (Z^{12})$ does not vanish. 
The middle cohomology of $X_\C^4$ splits as  
 $H^4(X^4_\C)=B_+(X^4_\C) \oplus B_{-}(X^4_\C)$
 into a selfdual ($*\omega=\omega$)
subspace $B_+(X^4_\C)$ and anti-selfdual  
($*\omega=-\omega$)
subspace $B_{-}(X^4_\C)$, whose dimensions are determined by the Hirzebruch
signature as
\bea
\sigma (X^4_\C)&=& \dim B_+(X^4_\C) - \dim B_{-}(X^4_\C)
\nonumber\\
&=& \int_{X^4_\C} L_2= \frac{1}{45}\int_{X^4_\C} (7p_2 - p_1^2)
= \frac{\chi (X^4_\C)}{3} + 32\;.
\label{tau formula}
\eea 
The symmetric inner product $(\omega_1, \omega_2)=\int_{X^4_\C} \omega_1 \wedge * \omega_2$ is positive 
definite on $H^4(X^4_\C)$, and $H^4(X^4_\C; \Z)$
is unimodular by Poincar\'e
duality. 
The symmetric quadratic form $Q(\omega_1, \omega_2)=\int_X \omega_1 \wedge \omega_2$
is positive definite on $B_+(X^4_\C)$ and negative definite on $B_{-}(X^4_\C)$. 
The reduction of the M-theory action is performed in \cite{Y} where 
the one-loop degree eight polynomial  $I_8$ is taken to have components 
along the Calabi-Yau space, which leads to a quantization on the values of the 
 Euler characteristic 
\(
I=-\int_{X^4_\C} I_8 = 
\int_{X^4_\C} (4p_2 - p_1^2)/192=\frac{\chi (X_\C^4)}{24} \in \Z\;.
\)
The C-field in M-theory satisfies
$[G_4] - \frac{1}{4}p_1 \in H^4(X^4_\C;\Z)$ \cite{Flux}.
If $G_4$ is zero then $\frac{1}{4}p_1$ has to be an integral class.
 This implies by Wu's formula that
$x^2 \equiv 0$ mod 2 for any $x \in H^4(X^4_\C;\Z)$.
From \eqref{tau formula}, this means that $H^4(X^4_\C;\Z)$ is an even
self-dual lattice with signature $\sigma (X^4_\C)$. The requirement that 
$\chi =0$ mod 24 is consistent with the fact that every even self-dual lattice
should have $\sigma =0$ mod 8. 
If $\frac{1}{4}p_1$ is half-integral then $[G_4]$ has to be half-integral,
and a potentially non-integral contribution to the one-point function
can be cancelled also for Calabi-Yau's for which 
$\chi \neq 0$ mod 24 \cite{Y}. 

\subsection{The signature (operator) in eleven dimensions and the C-field}

Now suppose $Z^{12}$ has a boundary $Y^{11}$ and is isometric to 
a product near the boundary. 
\footnote{The general case will be considered in section \ref{Sec s1}.}
Then near $Y^{11}$, $\sigma$ is of the form $\sigma=\left(\frac{\partial}{\partial z} +\cS \right)$
with $\cS$ a self-adjoint operator on $Y^{11}$. The restriction of $\Omega^+_Z$ to 
$Y^{11}$ can be identified with the space 
$\Omega^*_Y$
of all differential forms on $Y^{11}$.
On $Y^{11}$, the signature operator is defined as \cite{APS1}
\(
\cS \phi= (-1)^p (\epsilon * d - d*) \phi\;,
\)
where $*$ is the Hodge star operator defined by the metric $g_Y$
and with
 $\epsilon=1$ for $\phi$ a $2p$-form and 
$\epsilon=-1$ for $\phi$ a $(2p-1)$-form.  
The operator $\cS$ commutes with the parity of differential forms
on $Y^{11}$ and commutes with the operator which is essentially
 the Hodge duality operator $\phi \mapsto (-1)^p * \phi$.
When $p$ is even then  
 $\cS$ commutes precisely with $*$, and when $p$ is odd then 
 $\cS$ commutes with $*$ up to a sign. 
 This splits the operator $\cS$, 
 according to form degree,
  into even and odd parts
$\cS=\cS^{\rm ev} \oplus \cS^{\rm odd}$.

\paragraph{The even and odd signature operators.}
Let $(Y^{11}, g_Y)$ be a compact oriented eleven-dimensional 
Riemannian manifold. The odd signature operator $\cS^{\rm odd}_p$ is
defined on  $\bigoplus_{p=1}^6 \Omega^{2p-1}$, the differential forms 
of odd degree, by
\(
\cS^{\rm odd}_p=(-1)^{p+1}(*d + d*)\;.
\)
Similarly, the even signature operator 
$\cS^{\rm ev}_p$ is
defined on  $\bigoplus_{p=0}^5 \Omega^{2p}$, the differential forms 
of even degree, by
\(
\cS^{\rm ev}_p=(-1)^{p}(*d - d*)\;,
\)
The operator $\cS^{\rm ev}$ isomorphic to the operator
$\cS^{\rm odd}$ \cite{AS3}.

\paragraph{The even signature operator on the fields $G_4$ and $G_8$.}
We now consider the field $G_4$ and its Hodge dual $G_7$
in eleven dimensions out of which we build a field $G_8$. 
The equation of motion for $G_4$, 
\(
d*G_4=\frac{1}{2}G_4 \wedge G_4 - I_8(g_Y)\;,
\label{eom}
\)
involves $dG_7$, which we call 
$G_8$ (this is called $\Theta$ in \cite{DFM}).
The even signature operator 
$\cS (g_Y)=\oplus_p S_p : \Omega^{2*} (Y^{11}) \longrightarrow 
\Omega^{2*} (Y^{11})
$ 
on  $(Y^{11}, g_Y)$
acts on even degree forms as
\begin{eqnarray}
\cS_p : \Omega^{2p} (Y^{11}) &\longrightarrow &
\Omega^{10-2p} (Y^{11}) \oplus \Omega^{12-2p}(Y^{11})
\nonumber\\
G_{2p} &\longmapsto & (-1)^p (*d - d*) G_{2p}\;.
\label{sppp}
\end{eqnarray}
For $p=2$ this gives the action on the field strength $G_4$ of the C-field
 \begin{eqnarray}
\cS_2 : \Omega^{4} (Y^{11}) &\longrightarrow &
\Omega^{6} (Y^{11}) \oplus \Omega^{8}(Y^{11})
\nonumber\\
G_4 &\longmapsto &  *dG_4  - d*G_4\;,
\label{s2}
\end{eqnarray}
which, upon use of the Bianchi identity $dG_4=0$, gives
 $G_4 \mapsto - d*G_4$. This can further be expanded 
 using the equation of motion 
 \eqref{eom},
 resulting in 
 \(
 \cS_2: G_4 \longmapsto I_8(g_Y) - \frac{1}{2}G_4 \wedge G_4\;.
\label{s22}
 \)
 Next, for $p=4$, the operator \eqref{sppp} acting on the field strength $G_8$ gives
\(
\cS_4: G_8 \longmapsto * dG_8 - d* G_8\;.
\label{s4}
\)
As mentioned above, we 
take $G_8$ to be the right hand side of the equation of motion of the 
C-field, i.e.
$
G_8=d* G_4 - \frac{1}{2}G_4 \wedge G_4$.
Consequently, $dG_8=0$ and 
$
d* G_8= \Delta G_4 - \frac{1}{2}d* (G_4 \wedge G_4)$, so that 
\(
\cS_4: G_8 \longmapsto d* G_8= -\Delta G_4 + \frac{1}{2}d* (G_4 \wedge G_4)\;.
\label{s44}
\)
Here $\Delta=(d+d^*)^2$ is the Hodge Laplacian, which on $G_4$ and $G_8$
is simply $dd^*$ since both of these fields are closed.

\paragraph{The index  of $\cS_2$.}Consider the kernel of the operator 
\eqref{s2}. From \eqref{s22}, this space is
\(
{\rm Ker}(\cS_2)= \{G_4 ~|~ d*G_4=0\}
\)
so that it is given by requiring the eight-form on the 
right-hand side of the equation of motion \eqref{eom} to be zero
\(
\frac{1}{2}G_4 \wedge G_4 - I_8(g_Y)=0\;.
\)
This expression can be easily arranged to hold
 by requiring each of the two terms to vanish separately. 
 The quadratic term would be zero if $G_4$ is taken to have specific components 
which run over less than eight values. The one-loop term is zero for manifolds $(M, g_M)$ 
for which the Pontrjagin forms satisfy 
$p_2(g_M) - \frac{1}{4}p_1(g_M)^2=0$. At the level of cohomology
this is satisfied for eight-manifolds with nowhere vanishing spinors (see \cite{IPW})
or with a PU(3) structure (see \cite{W}).
The cokernel of the operator \eqref{s2} is the kernel of the 
adjoint operator 
$
\cS_2^\dagger
$
so that 
${\rm Coker}(\cS_2) = {\rm Ker}(\cS_2)^\dagger$ is given by 
$G_4=0$, that is by flat C-fields.
The dimension of the space of C-fields satisfying the equations of 
motion, when $G_4 \wedge G_4 - I_8 (g_Y)=0$,
is given by dim(Ker($\cS_2$)). 
The dimension of the space of flat C-fields is then given by 
dim(Ker($\cS_2^\dagger$)).
The index of the operator $S_2$ is  
\bea
{\rm Index}(\cS_2)&=& {\rm dim}({\rm Ker}(\cS_2)) 
- {\rm dim} ({\rm Coker}(\cS_2))
\nonumber\\
&=&\{ {\rm ``on-shell"~C-fields}\} - \{ {\rm flat~C-fields } \}\;.
\eea


\paragraph{The index  of $\cS_4$.}
Now consider the kernel of the operator \eqref{s4} acting on the dual field $G_8$,
$
{\rm Ker}(\cS_4)= \{G_8 ~|~ d*G_8=0\}$.
From \eqref{s44}, this is given in terms of $G_4$ by 
\(
{\rm Ker}(\cS_4)= \{G_8 ~|~ \Delta G_4 -\frac{1}{2}d* (G_4 \wedge G_4)=0\}\;.
\label{kers4}
\)
The
differential equation giving the condition in \eqref{kers4} 
 does not seem to have a general solution. Instead, we will give a characterization
in certain special cases. When $G_4 \wedge G_4=0$, as discussed above for 
 Ker($\cS_2$), the condition in \eqref{kers4} is simply that $G_4$ is harmonic. 
However, when $*(G_4 \wedge G_4)$ is a nonzero three-form, say proportional 
 to the C-field itself, with $dC_3=G_4$, then the condition is $(\Delta - m) G_4=0$, 
 that is $G_4$ is annihilated by the ``massive" Laplacian. Here $m$ is the parameter of 
 proportionality, that is $*(G_4 \wedge G_4)=2mC_3$. This is reminiscent of  the
 phenomenon of  odd-dimensional self-duality which appears in supergravity theories
 in odd dimensions, where a field strength is (Hodge) dual to a potential. 
 The cokernel of the operator $\cS_4$ is the kernel of the adjoint operator
$\cS_4^\dagger$, where
$G_8$ being zero implies that $G_7$ is constant, which is the same as 
$*G_4$ being a constant. The index can be found similarly to the case of 
$\cS_2$ (within the above specialization).

\paragraph{The odd  signature operator on the potential fields $C_3$ and $G_7$.}
We now consider the signature operator acting on the odd forms in M-theory.
These are the C-field $C_3$ and the field $G_7$, the Hodge dual of $G_4$. 
For the first field we have
\(
\cS_2^{\rm odd}: C_3 \longmapsto -(d* + *d)C_3= -G_7\;,
\)
 so that the kernel is given by flat dual fields. 
 For the second field we have
 \(
 \cS_4^{\rm odd}: G_7 \longmapsto -(d* + *d) G_7= -*d*G_4\;,
 \)
 so that the kernel is given by co-closed $G_4$. 
 The cokernels are given by zero $C_3$ and by flat C-fields, i.e. those for which
 $G_4=0$. Therefore, the indices are given, respectively, by 
 \(
 {\rm Index}(\cS_2^{\rm odd})= \{ {\rm dual~flat~fields}\} - \{ {\rm zero~ C-fields}\}
 \)
 \(
 {\rm Index}(\cS_4^{\rm odd})=
 \{ {\rm coclosed~} G_4  \} - \{  {\rm flat~ C-fields}  \}\;.
 \)
 

\paragraph{Alternative form.} Alternatively, instead of the operator 
$d* \pm *d$, we could use the operator 
$d \pm *d*$, which is manifestly (anti) self-adjoint. 
Then, for example, on $G_4$ we would get
$
G_4 \longmapsto *\left( \frac{1}{2}G_4 \wedge G_4- I_8(g)\right)$.



\paragraph{The eta-invariant.}
Since $\cS$ is self-adjoint and its square $\cS^2=\Delta$ is the Hodge Laplacian, it
is diagonalizable with real eigenvalues $\lambda_n$.
The eta-invariant $\eta (\cS(Y^{11}))$, which is a measure of asymmetry of the
spectrum of the operator, 
 is defined as the value at $s=0$ of \cite{APS1} 
\(
\eta (s)=\sum_{\lambda_n\neq 0} \frac{{\rm sign}\lambda_n}{|\lambda_n|^s}\;.
\)
The fact that the operator $\cS^{\rm ev}$ isomorphic to 
the operator $\cS^{\rm odd}$ implies, in particular, 
that the eta-invariant and number of zero modes 
corresponding to $\cS$ can be written in terms of those of $\cS^{\rm ev}$ 
as 
\(
\eta (\cS)=2 \eta (\cS^{\rm ev})\;, \qquad \qquad h(\cS)=2h(\cS^{\rm ev})\;,
\)
and similarly for $\cS^{\rm odd}$.
Therefore, in dealing with the eta-invariant, one can formulate 
expressions using the `total' signature operator, the odd signature
operator, or the even signature operator, with the simple prescribed way 
of transforming from one formulation to the other.

\vspace{3mm}
Having discussed the effect of the signature operator on the fields, next we turn to 
the corresponding effect on the action and partition function of the theory.

\section{The invariants from M-theory}
\label{sec inv}
In this section we start by recalling the APS index theorem for the 
signature operator and then provide our main arguments in this context in 
section \ref{sec q} and section \ref{sec dif}.

\paragraph{The APS index theorem for the signature operator.}
Let ($Z^{12}$, $g_Z$)
 be a compact oriented Riemannian manifold 
with boundary ($Y^{11}$, $g_Y$),
 and assume that near the boundary the twelve-manifold
  is isometric to a product. Then the APS index theorem relates a topological invariant 
on one side to a sum of a differential geometric and a spectral
invariant on the other side \cite{APS1}
\(
{\rm sign}(Z^{12}, Y^{11}) = 
\int_{Z^{12}} L(p(g_Z)) - \eta (\cS(g_Y))\;,
\label{aps sig}
\)
where 

\noindent {\bf (i)} sign$(Z^{12})$ is the signature of the nondegenerate quadratic 
form defined by the cup product on the image of 
$H^6(Z^{12}, Y^{11})$ in $H^6(Z^{12})$. Looking at relative cohomology is 
appropriate since there are no six-form field strengths in M-theory in eleven dimensions.

\noindent {\bf (ii)} $L(p(g_Z))=L_{12}(p_1, p_2, p_3)$, where $L_{12}$ is the 
3rd Hirzebruch L-polynomial (of degree twelve) and the $p_i$ are the Pontrjagin 
forms of the curvature built out of the Riemannian metric $g_Z$. 

\noindent {\bf (iii)} $\eta (\cS(g_Y))$ is the eta-function for the self-adjoint operator 
$\cS_p$ on even forms on $Y^{11}$ given by 
$G_{2p} \mapsto (-1)^p (*d-d*)G_{2p}$. The multiplicity of the zero eigenvalue of 
$\cS_p$
is $h={\rm dim}({\rm Ker}(\cS_p$)). For the values $p=2, 4$,
 we considered this in the previous section.

\vspace{3mm}
Notice that 
if sign$(Z^{12}, Y^{11})=0$ then the topological quantity is given in terms of the 
geometric/analytical quantity
$\int_{Z^{12}} L(p(g_Z))=\eta (\cS(g_Y))$. 
This can happen, for instance, when $Z^{12}$ itself
is a boundary, as indicated in the bordism property mentioned in 
section \ref{sig 12}.

\subsection{A variation on the miraculous cancellation formula in twelve dimensions}
\label{sec q}
The topological part of the action in M-theory,
namely the combination of the Chern-Simons term and the one-loop 
term 
\(
I=I_{CS} + I_{\rm 1-loop}= \frac{1}{6}\int_{Z^{12}} G_4 \wedge G_4 \wedge G_4
-\int_{Z^{12}} I_8 \wedge G_4\;,
\label{top}
\)
is formulated in \cite{Flux} in terms of index theory.
This involves $I^{E_8}$, the index of the Dirac operator coupled to an $E_8$ bundle, 
as well as $I^{RS}$, the index of the Rarita-Schwinger operator, that is the 
Dirac operator coupled to the virtual vector bundle $TZ^{12} - 4 \O$.
The subtraction of four copies of the trivial line bundle, $-4 \O$, 
comes from the inclusion of effect of ghosts    
required to fix the gauge invariances of the Rarita-Schwinger operator
 in eleven dimensions.
The exponent in the phase of the partition function is \cite{Flux}
\(
\frac{1}{2\pi} I= \frac{1}{2}I^{E_8} + \frac{1}{4}I^{RS}\;.
\label{phase}
\)
The factor of $1/2$ on the right hand side is due to a Mojorana-Weyl (MW) condition.
The factor of $1/4$ is due to a MW condition and the fact that the
characteristic class
should to be related to half of the gravitino-dilatino anomaly (in comparing to 
heterotic string theory).

\vspace{3mm}
We now provide our alternative description, using the the Hirzebruch signature 
theorem, and hence the Hirzebruch L-polynomial. 
We start with  the Rarita-Schwinger index
in twelve dimensions. This is given by 
\bea
I^{RS}_{12} &=& \widehat{A}(Z^{12}) {\rm ch} (TZ^{12}_\C - 4\mathcal{O})
\nonumber\\
&=&
\frac{1}{2^{10} \cdot 3^3 \cdot 5 \cdot 7}
\left(
-31 p_1^3 + 44 p_1p_2 -16 p_3
\right) \cdot 8
+ 
\frac{1}{2^7 \cdot 3^2 \cdot 5}\left(7 p_1^2 -4p_2 \right)\cdot p_1
\nonumber\\
&&- \frac{1}{2^3 \cdot 3}(p_1) \cdot \frac{1}{2^2 \cdot 3}
(p_1^2 - 2p_2)
+ 1 \cdot \frac{1}{2^3 \cdot 3^2 \cdot 5}
\left( 
p_1^3 - 3p_1 p_2 + 3p_3
\right)
\nonumber\\
&=& \frac{1}{2^3 \cdot 3^3 \cdot 5 \cdot 7}
\left(2p_1^3 -13 p_1 p_2 + 62p_3 \right)
\eea
On the other hand, the degree-twelve part of the Hirzebruch L-polynomial
is given by (see e.g. \cite{Hir})
\(
L_{12}= \frac{1}{3^3 \cdot 5 \cdot 7}
\left( 
62p_3 -13p_1p_2 + 2p_1^3
\right)\;.
\)
Now we get a formula which is
a variation on the miraculous anomaly cancellation formula 
of Alvarez-Gaum\'e
and Witten \cite{AW}. This relies on the curious degree twelve expression 
\footnote{We will label the $\widehat{A}$-genus and the L-genus by their 
form-degree rather than by that divided by four.} 
\(
8 I^{RS}_{12}= L_{12}\;.
\label{trade}
\)
To some extent, our formula can be viewed as a quantum counterpart of 
 the classical miraculous cancellation formula 
By ``quantum" we mean in the sense of accounting for ghosts
coming from the path integral are
accounted for. However, a fully quantum version would involve setting up 
effective actions as in \cite{FM}, which will be discussed separately elsewhere.
Note that for a twelve-manifold $M$ with tangent bundle $TM$ the miraculous cancellation formula is \cite{AW}
\(
L(M) = 8 \widehat{A}(M, TM) - 32 \widehat{A}(M)\;,
\)
where $\widehat{A}(M, TM)=\widehat{A}(M) {\rm ch}(TM)$, with
${\rm ch}(TM)=\sum_j e^{x_j} + e^{-x_j}=\sum_j 2 {\rm cosh} x_j$. 
It is easily seen that
$
L(M)=8 \widehat{A} (M) [ {\rm ch}(TM) - 4]
$.
We now rewrite \eqref{phase}, the exponent in the phase of the partition function, 
arriving at the alternative expression which trades the Rarita-Schwinger 
index with the Hirzebruch L-polynomial via \eqref{trade}
\(
\frac{1}{2\pi}I=\frac{1}{2}I^{E_8} + \frac{1}{32}L_{12}\;.
\label{12I}
\)
To motivate what might be gained by doing this,
let us consider the case when the $E_8$ bundle is trivial. In this case the
top degree component of \eqref{12I} reduces to
\(
\frac{1}{2\pi}I= 124 \left( \widehat{A}_{12} + \frac{1}{2^7\cdot 31} L_{12}  \right)\;.
\label{eq al12}
\)
Absence of anomalies requires that the right hand side be
 an integer (as it is a phase in the effective action), 
 so that we arrive at the condition
\(
\left\langle \widehat{A}_{12} + \frac{1}{2^7\cdot 31} L_{12}, [Z^{12}] \right\rangle \in \Z/124\;.  
\label{no e8}
\)

\begin{proposition} (i). The miraculous cancellation formula in twelve dimensions can be written as
$8I_{12}^{RS}=L_{12}$, where $I_{12}^{RS}$ is the Rarita-Schwinger index and $L_{12}$ is 
the Hirzebruch L-polynomial in degree 12. 

\noindent (ii). The topological action in M-theory is $\frac{1}{2}I^{E_8} + \frac{1}{32}L_{12}$, where
$I^{E_8}$ is the index of the Dirac operator coupled to an $E_8$ bundle. 

\noindent (iii). The phase is not anomalous if 
$\left\langle \widehat{A}_{12} + \frac{1}{2^7\cdot 31} L_{12}, [Z^{12}] \right\rangle \in \Z/124$.
\end{proposition}

In the following section we show that this naturally leads to the Kreck-Stolz 
$s$-invariant.  

\subsection{The Kreck-Stolz $s$-invariant and
 scalar curvature in eleven dimensions}
\label{sec dif}
Let $Z^{12}$ be a twelve-dimensional compact Spin manifold with boundary
$\partial Z^{12}=Y^{11}$. 
Let $g_Z$ be a Riemannian metric on $Z^{12}$
which coincides with
a product metric on $Y^{11} \times I$ in a collar neighborhood of the 
boundary and let $g_Y$ be its restriction to the boundary. 
Let $D^+(Z^{12}, g_Z)$ be the (chiral) Dirac operator with respect to the 
metric $g_Z$ from the positive to negative 
spinors on $Z^{12}$. This becomes a Fredholm operator if we impose the 
APS boundary condition \cite{APS1}, i.e. if we restrict to spinors on $Z^{12}$ 
whose restriction to $\partial Z^{12}$ is in the kernel of $P$,
the spectral projection corresponding to nonnegative 
eigenvalues of the (total) Dirac operator $D(Y^{11}, g_Y)$ on $Y^{11}$.
Denote by ${\rm index}(D^+(Z^{12}, g_Z))$ the index of this Fredholm operator.
If $g_{Z}(t)$ is a continuous family of metrics on $Z^{12}$, then 
this index is independent of $t$, which can be seen as follows 
(cf. \cite{APS1} \cite{KS}).
The corresponding family of spectral projections $P(t)$
is not continuous for those parameter
values $t$ where an eigenvalue of $D(Y^{11}, g_Y)$ crosses the origin. 
If $g_Y(t)$ has positive scalar curvature metric then 
the Weitzenb\"ock formula gives \cite{Li} that 
Ker$(D(Y^{11}, g_Y(t)))=0$. Hence $D^+(Z^{12}, g_Z(t))$ is a
continuous family of Fredholm operators and thus 
${\rm index}(D^+(Z^{12}, g_Z(t)))$ is independent of $t$. Note the following:

\begin{enumerate}
\item If $g_Z$ has a positive scalar curvature metric then,
from \cite{APS2}, ${\rm index}(D^+(Z^{12}, g_Z))$ vanishes. 

\item If $g_Z$ is a metric on $Z^{12}$ whose restriction to the boundary
$g_Y$ has positive scalar curvature then ${\rm index}(D^+(Z^{12}, g_Z))$  depends only
on the connected component of $g_Y$ in $\mathcal{R}^+ (Y^{11})$,  
the space of positive scalar curvature metrics on $Y^{11}$ \cite{KS}.
\end{enumerate}

\noindent The APS index theorem for the Dirac operator is \cite{APS1}
\(
{\rm index}(D^+(Z^{12}, g_Z)) = \int_{Z^{12}} \widehat{A} (p_i( g_Z)) -
\frac{1}{2}\left( h(Y^{11}) + \eta (D(Y^{11}, g_Y) \right)\;.
\label{aps d}       
\)
Here 

\noindent -- $p_i(g_Z)$ are the Pontrjagin forms of $Z^{12}$
with respect to the Levi-Civita connection  $\nabla^L_Z$ 
determined by $g_Z$,

\noindent -- $D(Y^{11}, g_Y)$ is the Dirac operator on $Y^{11}=\partial Z^{12}$,

\noindent -- $h(Y^{11})$ is the dimension of the kernel of $D(Y^{11}, g_Y)$
which consists of harmonic spinors on the boundary,

\noindent -- and $\eta (D(Y^{11}, g_Y))$ is the $\eta$-invariant, which measure 
the asymmetry of the spectrum of the self-adjoint operator $D(Y^{11}, g_Y)$.

\paragraph{Additivity and the space of harmonic spinors on $Y^{11}$.}
Consider two twelve-manifolds $(Z^{12}, g_Z)$ and $(Z'^{12}, g_Z')$.
If we glue these two manifolds along isometric boundary
component $Y^{11}$, then the extension of the Chern-Simons term
$I_{CS}=\frac{1}{6} \int_{Y^{11}}G_4 \wedge G_4\wedge C_3$ from 
$Y^{11}$ to $Z^{12}$ leading to $I_{CS}=\frac{1}{6}\int_{Z^{12}} G_4 \wedge G_4 \wedge G_4$,
is independent of the choice of the bounding twelve-manifold \cite{Flux}.
However, 
 the index formula shows that the 
index of the Dirac operator behaves additively, provided that there are no 
harmonic spinors on $Y^{11}$, i.e. $h(Y^{11})=0$. This happens, for example, 
if the scalar curvature of that piece of the eleven-dimenisonal 
boundary is positive. Such situations are
considered extensively in \cite{DMW-Spinc}.

\paragraph{The Kreck-Stolz invariant $s(Y^{11}, g_Y)$.}
We will arrive at 
an invariant $s (Y^{11}, g_Y) \in \Q$, defined in \cite{KS}, 
as an absolute version of the 
Gromov-Lawson invariant \cite{GL}, which in our case would be
 for a pair of positive scalar curvature
metrics $g_1$ and $g_2$ on $Y^{11}$. 
We will first describe this invariant, following \cite{KS}, and then show
how M-theory leads to it naturally.
This invariant is obtained by rewriting \eqref{aps d} as a sum of two terms,
one depending only on the geometry of $Y^{11}$ and another depending only
on the topology of $Z^{12}$. This can be done provided that the 
real Pontrjagin classes of $Y^{11}$ vanish. 
The construction relies on treating the 
decomposable vs.
nondecomposable summands in $\int_{Z^{12}} \widehat{A}(p_i(g_Z))$
separately \cite{KS}.

\vspace{2mm}
\noindent  {\it Decomposable summands:} Let $\alpha_4$ and $\beta_8$ be differential 
forms of positive degree on $Z^{12}$ whose restrictions to the boundary
 $Y^{11}$ are coboundaries, i.e. there are forms 
 $c_3$ and $c_7$ on $Y^{11}$ such that $dc_3=\alpha_4|_{Y^{11}}$
 and $dc_7= \beta_8|_{Y^{11}}$. Then the wedge products are related as
 \(
 \int_{Z^{12}} \alpha_4 \wedge \beta_8 = \int_{Y^{11}} \alpha_4 \wedge c_7 
 + 
 \left\langle 
 j^{-1}[\alpha_4] \cup j^{-1} [\beta_8]~ ,~ [Z^{12}, Y^{11}]
 \right\rangle \;,
 \label{jgg}
 \)
where $j^{-1}[\alpha_4] \in H^4(Z^{12}, Y^{11}; \R)$ in any 
preimage of the de Rham cohomology class $[\alpha_4]\in H^4(Z^{12};\R)$ under
the natural map $j: H^4(Z^{12}, Y^{11};\R) 
\to H^4(Z^{12};\R)$ and similarly for $[\beta_8]$, and 
$\langle ~~, [Z^{12}, Y^{11}]\rangle$ is the Kronecker product with the
fundamental class. 
Note that the integral on the right hand side 
of \eqref{jgg} is independent of the 
choice of $c_7$ and the Kronecker product is independent 
of the preimages $j^{-1}[\alpha_4]$ and $j^{-1}[\beta_8]$.
Taking $\alpha_4$ and $\beta_8$ to be rational multiples of 
the Pointrjagin forms $p_1( g_Z)$ and $p_2( g_Z)$, 
respectively, gives that the decomposable summands in 
$\widehat{A}(p_i(g_Z))$ can be written as 
a sum of two terms.

\vspace{2mm}
\noindent {\it Nondecomposable summands:} 
This summand in $\widehat{A}(p_i(g_Z))$
is a nontrivial multiple of the top 
Pontrjagin form $p_3(g_Z)$. 
Since the Hirzebruch L-polynomial also involves
this form (with another multiple) then one can 
arrange for a combination 
of $\widehat{A}$ and L which cancels $p_3$, namely
\cite{Hir} the following combination $\widehat{A}_{3} + \frac{1}{2^7\cdot 31}L_{3}$. 
Let $j^{-1}p_i(Z^{12})$ be any preimage 
under the natural map $j: H^{4i}(Z^{12}, Y^{11};\R) 
\to H^{4i}(Z^{12};\R)$. This exists because we are 
assuming $p_i(Y^{11})=0\in H^{4i}(Y^{11};\R)$.
 Then
\bea
{\rm index}(D^+ (Z^{12}, g_Z))&=&
\int_{{}_{Y^{11}}} d^{-1}\left( \widehat{A} + \frac{1}{2^7\cdot 31}L\right)
(p_i(Y^{11}, g_Y))
-\frac{1}{2}\left( h(Y^{11}) + \eta (D(Y^{11}, g_Z|_Y) \right)
\nonumber\\
&&- \frac{1}{2^7\cdot 31} \eta(\cS(Y^{11}, g_Z|_Y)) - t(Z^{12})\;,
\eea
where the topological term is 
\(
t(Z^{12})= - \left\langle (\widehat{A} + \frac{1}{2^7\cdot 31}L) (j^{-1}p_i(Z^{12})),
[Z^{12}, Y^{11}]
\right\rangle +  \frac{1}{2^7\cdot 31} {\rm sign}(Z^{12})\;.
\)
In particular, if all the real Pontrjagin classes 
of $Y^{11}$ vanish then we can apply the formula to
$Z^{12}= Y^{11}\times I$, in  which case $t(Z^{12})$ vanishes. 

\vspace{3mm}
Given a closed eleven-dimensional Spin manifold
$Y^{11}$ 
with vanishing real Pontrjagin classes and positive
scalar curvature metric $g_Y$ on $Y^{11}$ we define,
following \cite{KS}, 
\(
s(Y^{11}, g_Y):=-\frac{1}{2}\eta(D(Y^{11}, g_Y)) -
 \frac{1}{2^7\cdot 31}\eta (\cS(Y^{11}, g_Y)) +
 \int_{Y^{11}} d^{-1} \left(\widehat{A}_{12} +  \frac{1}{2^7\cdot 31} L_{12}\right) (p_i(Y^{11}, g_Y))\;.
\label{exp ks}
\)

\paragraph{Properties of $s(Y^{11}, g_Y)$.}
Let $Y^{11}$ and $Y'^{11}$ be eleven-dimensional closed Spin manifolds
with vanishing real Pontrjagin classes and 
positive scalar curvature metrics $g_Y$ and $g'_Y$, respectively. Then, 
specializing \cite{KS}, 
\begin{enumerate}
\item If $f: s(Y^{11}, g_Y) \to s(Y'^{11}, g'_Y)$ is a 
Spin preserving isometry, then 
$s(Y^{11}, g_Y)= s(Y'^{11}, g'_Y)$.
\item $s(Y^{11}, g_Y)$ depends only on the connected component of 
$g_Y$ in $\mathcal{R}^+_{\rm scal}(Y^{11})$, the moduli space of positive 
scalar curvature metrics on Spin eleven-manifolds.
\item If $Y^{11}$ bounds a Spin manifiold $Z^{12}$ and $g_Z$ is a metric 
on $Z^{12}$ extending $g_Y$, which is a product metric near the boundary, then
\(
s(Y^{11}, g_Y)={\rm index}(D^+ (Z^{12}, g_Z)) + t(Z^{12})\;.
\)
\item $s(Y^{11}, g_Y)$ depends on the choice of Spin structure on $Y^{11}$.
\end{enumerate}

Consider the expression \eqref{eq al12}
 on a twelve-manifold with boundary. Using the APS index theorem, both 
 for the signature operator \eqref{aps sig}
and for the Dirac operator  \eqref{aps d}, we get that the phase of the M-theory partition 
function in eleven dimensions is given by expression  \eqref{exp ks}.
 Given the  identification of the phase in the M-theory partition function essentially 
 with the  Kreck-Stolz $s$-invariant, the anomaly cancellation condition \eqref{no e8}
in twelve dimensions
 can now be recast as saying that in eleven dimensions $s(Y^{11}, g_Y) \in \Z/124$. 
We therefore have

\begin{theorem}Consider M-theory on Spin $(Y^{11}, g_Y)$, where
$g_Y$ is a metric of positive scalar curvature, and let the $E_8$ bundle 
on $Y^{11}$ be trivial. Then

\noindent (i). The phase of the M-theory partition function is anomaly free provided 
$s(Y^{11}, g_Y) \in \Z/124$.

\noindent (ii). M-theory on a Spin manifold with positive scalar curvature metric
$(Y^{11}, g_Y)$ detects diffeomorphism types. 

\noindent (iii). The topological part of the action is invariant under Spin isometries. 

\noindent (iv). The topological part of the action depends only on the connected component 
of the metric in the moduli space of positive scalar curvature metrics. 

\noindent (v). The topological part of the action depends on the choice of Spin structure. 
\label{big thm}
\end{theorem}
The last part of the theorem is discussed extensively in \cite{DMW-Spinc} from 
another point of view. See also section \ref{Sec s1}.

 \paragraph{Conditions and examples of $s$-invariants satisfying anomaly cancellation.}
 We would like to check that the condition \eqref{no e8} or, more precisely the condition in 
 part $(i)$ of Theorem \ref{big thm}, is satisfied for some relevant Spin eleven-manifolds. 
 Since the eta-invariant is additive under direct sum, we could consider decomposable
 manifolds and restrict to 
 seven-manifolds, as internal spaces of compactifications to 
 four dimensions. An important class of such Spin Einstein 
 manifolds which solve the supergravity equations of motion 
 is the Witten spaces
 $M_{k,l}$ \cite{KK}, which are principal $S^1$ bundles over $\C P^2 \times \C P^1$ 
 classified by $l x + k y \in H^2(\C P^2 \times \C P^1;\Z)$, where $x$ and 
 $y$ are the generators of $H^2(\C P^2;\Z)$ and 
 $H^2(\C P^1;\Z)$, respectively \cite{KS0}.
  Let $k$ and $l$ be relatively prime integers with $k$ even. Then the $s$-invariant 
  of $M_{k, l}$ with Einstein metric $g_{k, \ell}$ is \cite{KS}
  \(
  s(M_{k, l}~,~ g_{k, l})=-\frac{3}{2^7 \cdot 7}\frac{k(l^2 +3) (l^2-1)}{l^2}\;,
  \) 
  from which we observe that for
  the values $(k, l)=(14,3)$ we get $124s(M_{14,3}~,~ g_{14,3})=744$.
  With this integer value for $s$, there are no anomalies in the phase. 
  Another example we consider is the family of 
  Aloff-Wallach spaces of positive sectional curvature,
  for which the $s$-invariant is \cite{KS}
  \(
  S(N_{k,l}~,~ g_{k,l})= \frac{1}{2^5\cdot 7}{kl(k+l)}\;.
  \)
  We see that for the values $(k,l)=(8,7)$ we find that $s(N_{8,7}~,~g_{8,7})=15/4$, so that
  the phase is $2\pi i$ times the integer $3\cdot 5\cdot 31$. 
  Other values can also be found by solving the above Diophantine equations
  (but we do not need that here).


\paragraph{The Eells-Kuiper invariant and stably parallelizable eleven-manifolds.}
We now consider a special class of eleven-manifolds, for which the $s$-invariant 
reduces to a more classical invariant. 
Let $Y^{11}$ be a stably parallelizable compact Spin 
eleven-manifold without boundary. That is, $TY^{11} \oplus \mathcal{O}$ is trivial.
Let $Y^{11}=\partial Z^{12}$ with $Z^{12}$ a Spin 
twelve-manifold. The Eells-Kuiper invariant is defined as \cite{EK}
\(
ek(Y^{11})= 
\left\langle
\left( \widehat{A}_{12}(p_1, p_2, p_3) - 
\frac{1}{2^7 \cdot 31} L_{12}(p_1, p_2, p_3)
 \right), [Z^{12}, Y^{11}] \right\rangle
+ 
\frac{1}{2^7\cdot 31} {\rm sign} (Z^{12}) \in \Q/\Z\;,
\)
where $p_i$ are the relative Pontrjagin classes of 
$(Z^{12}, Y^{11})$ corresponding to some framing of the 
stable tangent bundle of $Y^{11}$.
Let $\omega$ be a connection on the stable tangent bundle 
of $Y^{11}$. Extend $\omega$ over $Z^{12}$ as a product on a 
collar neighborhood. Following \cite{Do}, the Chern-Simons invariants are
given by
\bea
CS(L, Y^{11}, \omega)& =& {\rm sign} (Z^{12}) - \int_{Z^{12}} L_{12}(p_i(\omega))  \qquad \in \R\;,
\nonumber\\
CS(\widehat{A}, Y^{11}, \omega)&= &-\int_{Z^{12}} \widehat{A}_{12} (p_i(\omega))
\qquad \qquad \in \R/\Z\;.
\eea
If $\omega$ is the trivial 
connection 
corresponding to some framing then 
$p_i(\omega)$ represent the relative Pontrjagin classes of that 
framing and 
\(
ek({Y^{11}})= -CS(\widehat{A}, Y^{11}, \omega) + \frac{1}{2^7 \cdot 31} CS(L, Y^{11}, \omega) \qquad \in \Q/\Z\\;.
\)
Suppose $Y^{11}$ is endowed with a Riemannian metric $g_Y$. Define the
signature operator on even forms on $Y^{11}$ as above, and extend 
$g_Y$ to a metric on $Z^{12}$ as a product near the
boundary. Let $\nabla_{g_Y}$ be the Levi-Civita connection associated with $g_Y$. 
The APS index theorem \cite{APS1} gives the eta invariant for this metric
\bea
\eta (\cS(Y^{11})) &=& CS(L, Y^{11}, \nabla_{g_Y}) 
\\
\frac{1}{2} \left( h- \eta(D(Y^{11})) \right)
&=&
CS(\widehat{A}, Y^{11}, \nabla_{g_Y})) \qquad \in \R/ \Z\;.
\eea
 Let $f$ be any section of the stable tangent bundle and 
$TP(\omega)$ are the canonical forms satisfying $dTP(\omega)= P(\omega)$,
which is another way of writing $d^{-1}$.
Then, from \cite{Do}, 
\(
ek(Y^{11})= \frac{1}{2} \left( \eta(D(Y^{11})) -h \right)
+ \frac{1}{2^7 \cdot 31} \eta(\cS(Y^{11}))
-
\int f^* \left( T \widehat{A} (\nabla_{g_Y} + \nabla_0)
- \frac{1}{2^7 \cdot 31} TL (\nabla_{g_Y} - \nabla_0)
 \right)  \in \Q/\Z\;,
\)
where $\nabla_0$ is a trivial connection on $TY^{11}$. 
This is analogous to similar discussions on framing in \cite{tcu}. 
In this case of stably parallelizable manifolds, the phase 
of the partition function is given by the Eells-Kuiper invariant. 
Since $ek(Y^{11})$ classifies topological eleven-spheres, then 
\begin{observation}
The topological action in M-theory classifies topological eleven-spheres.
\end{observation}
This is related to the global gravitational anomalies of 
\cite{CMP} although M-theory is not chiral.

 \paragraph{A generalization of the Kreck-Stolz invariant?}
We define a new expression which, in addition to dependence on the metric and Spin structure, 
depends also on a degree four cohomology class $a$. Recall that in the M-theory 
expression \eqref{no e8}, which led to the Kreck-Stolz $s$-invariant, we assumed that 
the $E_8$ bundle in eleven and twelve dimensions is trivial, that is its degree four 
characteristic class $a$ is zero. Note that $BE_8 \sim K(\Z,4)$ in our range of dimensions so that
$a$ can take on any value.  
However, the action in M-theory involves an $E_8$
bundle which is in general not trivial. Therefore, we would like to consider the effect of 
including this class, together with the geometry. Assuming as in \cite{KS} that the 
real Pontrjagin classes vanish, implies in particular that the first Pontrjagin class 
appearing in the flux quantization condition of \cite{Flux}, $G_4 - \frac{1}{4}p_1=a \in H^4(Y^{11};\Z)$,
is absent so that allowing a degree four class is essentially the same as 
allowing a C-field through its field strength $G_4$, at least rationally away from torsion.
 The inclusion of the nontrivial class $a$ leads to a contribution of the corresponding
 Pontrjagin character of the $E_8$ bundle $E$, ${\rm Ph}(E)=248+
 60a + 6a^2 + \frac{1}{3}a^3$. Thus, we have 

\begin{definition}
$s(Y^{11}, g_Y, a)= s(Y^{11}, g_Y)+ d^{-1}\left( \frac{1}{3}a^3 + 6a^2 \widehat{A}_4
+ 60a \widehat{A}_2 \right)$.
\end{definition}
We propose this as the geometric invariant when a nontrivial C-field is 
present. 
While we have written this invariant in eleven dimensions (as relevant for M-theory),
the extension to other dimensions is obvious from our
use of the index theorem. It would be interesting to work this out explicitly.


\section{Comparison to type IIA string theory in ten dimensions}
In this section we relate the expressions we considered above 
in section \ref{sec inv},
for 
M-theory in eleven and (extension to) twelve dimensions, to 
corresponding ones in string theory in ten dimensions.
This comparison requires $Y^{11}$ to be a circle bundle. 


\subsection{The $s$-invariant for $Y^{11}$ a 
circle bundle}
\label{Sec s1}
Consider $Y^{11}$ to be the principal circle bundle $S^1 \to Y^{11} \buildrel{\pi}\over{\longrightarrow} X^{10}$ with positive scalar curvature metric $g_Y$, as considered in \cite{DMW-Spinc}.
Corresponding to the circle bundle is a complex line bundle $\cL$ with first Chern class
$c=c_1(\cL)$.
The tangent bundle splits as $TY^{11}\cong \pi^*(TX^{10}) \oplus T_FY^{11}$, 
where the tangent bundle along the fibers $T_FY^{11}$ trivial, with a trivialization 
provided by the vector field generating the $S^1$-action on $Y^{11}$.
Since we are assuming $p_i(Y^{11})=0\in H^{4i}(Y^{11};\R)$, $i=1,2$, 
the splitting of the tangent 
bundle implies that $\pi^*(p_i(X^{10}))=p_i(Y^{11})=0$. 
Then the Gysin exact sequence
\(
\xymatrix{
\cdots \ar[r] &
H^{4i-2}(X^{10};\R) 
\ar[r]^{\cup c} &
H^{4i}(X^{10};\R)
\ar[r]^{\pi^*}
&
H^{4i}(Y^{11};\R)
\ar[r] & \cdots
}
\)
relates the fields on $Y^{11}$ to the fields on $X^{10}$ (see \cite{MS}) and in our case
shows that $p_i(X^{10})$ is divisible by $c$. That is, there are elements 
(in the notation of \cite{KS})
$\overline{p}_i \in H^{4i-2}(X^{10};\R)$ such that $p_i(X^{10})= \overline{p}_ic$, $i=1,2$. 
This gives that the Pontrjagin classes of $X^{10}$ are zero when 
the Chern class of the line bundle is zero; otherwise they are in general
not zero.

\paragraph{The bilinear form.} As above,
consider a line bundle $\cL$ with Euler class $e=c(\cL)\in H^2(X^{10})$
over a base manifold $X^{10}$.
There is a natural symmetric bilinear form on the degree four cohomology 
\(
B_c: H^4(X^{10};\Z) \times H^4(X^{10};\Z) \to \R
\)
defined by $B_c (a, b) := \langle a \cup b \cup c(\cL) , [X^{10}] \rangle$, where 
$[X^{10}]$ is the fundamental homology class of $[X^{10}]$. 
This bilinear form is part of the expression for the phase in the Spin${}^c$ case
(see \cite{DMW-Spinc}).
The Thom 
isomorphism theorem implies that 
\(
{\rm sign}(\mathbb{D}(\cL))={\rm ~signature~of~}B_c\;,
\)
where $\pi_D: \mathbb{D}(\cL)=Y^{11} \times_{S^1} \mathbb{D}^2 \to X^{10}$ is the disk
bundle associated to the $S^1$-action on $Y^{11}$ with orbit manifold 
$X^{10}=Y^{11}/S^1$. 
In the general case when the Pontrjagin classes of $Y^{11}$ are not 
required to vanish, the obstruction to expressing the signature of the disk bundle 
$\mathbb{D}(\cL)$ as an evaluation of a characteristic class on $X^{10}$
is the limiting eta-invariant \cite{Ti}.

\paragraph{Dependence of the $s(Y^{11}, g_Y)$ on the Spin structure on $Y^{11}$.}
There are two cases to consider:
\begin{enumerate}
\item {\it $X^{10}$ is Spin}: In this case $w_2(X^{10})=0 \in H^2(X^{10};\Z_2)$.
By the decomposition of the tangent bundle of $Y^{11}$ we see that a Spin 
structure on $X^{10}$ induces a Spin structure on $Y^{11}$, which we denote
by $\xi$. 
\item {\it $X^{10}$ is Spin${}^c$}: In this case $w_2(X^{10})=c$ mod 2. 
Then $TX^{10}\oplus \cL$ 
admits a Spin structure. The choice of such a Spin structure gives
a Spin structure on the disk bundle $\mathbb{D}(\cL)$ whose restriction to the sphere bundle
$S\cL=Y^{11}$ is a Spin structure $\xi'$.
\end{enumerate}
In the Spin case, i.e. when $w_2(X^{10})=0$ and $c=0$ mod 2, we have that
$\xi$ and $\xi'$ are different Spin structures on $Y^{11}$, since the 
restriction of $\xi$ to a fiber $S^1$ is the nontrivial Spin structure,
which does not extend over the 2-disk $\mathbb{D}^2$, whereas the restriction of 
$\xi'$ extends by construction.
This is again discussed more fully in \cite{DMW-Spinc}.

\paragraph{The $s$-invariant for the case when $X^{10}$ is Spin${}^c$.}
The disk bundle $\mathbb{D}(\cL)$ is a twelve-manifold with 
boundary $Y^{11}$ and induced Spin structure $\xi'$ on that 
boundary. Then the index ${\rm index}( D^+(\mathbb{D}(\cL), g_{\mathbb{D}(\cL)}))=0$ for any metric
$g_{\mathbb{D}(\cL)}$ on ${\mathbb{D}(\cL)}$ which restricts to $g_Y$ on the boundary and 
is a product metric in a collar neighborhood of the boundary. 
Now for a line bundle $\cL$ of Chern class $c$, the genera are given by \cite{Hir}
\(
\widehat{A}(\cL)=\frac{c}{2{\rm sinh}(c/2)}\;, 
\qquad 
L(\cL)=\frac{c}{{\rm tanh}(c)}\;, 
\)
so that the s-invariant, using \cite{KS}, is
\(
s(Y^{11}, \xi', g_Y)=\left\langle
\widehat{A}(TX^{10})  \frac{1}{2{\rm sinh}(c/2)}
+\frac{1}{2^7\cdot 31} L(TX^{10}) \frac{1}{{\rm tanh}(c)},
[X^{10}]
\right\rangle
+ 
\frac{1}{2^7\cdot 31} {\rm sign} (B_c)
\)

\paragraph{The $s$-invariant for the case when $X^{10}$ is Spin.}
The general discussion is more difficult since there is no obvious 
Spin twelve-manifold $Z^{12}$ bounding $Y^{11}$ with the Spin
structure $\xi$.
It is also very difficult to compute the index ${\rm index}(D^+(Z^{12}, g_Z))$.
 However, as argued more generally in \cite{KS},
 when $g_Z$ has a metric of positive scalar 
curvature then the index is zero, in which case the $s$-invariant is given 
by
\(
s(Y^{11}, \xi, g_Y)=\left\langle
\widehat{A}(TX^{10})  \frac{1}{2{\rm tanh}(c/2)}
+\frac{1}{2^7\cdot 31} L(TX^{10}) \frac{1}{{\rm tanh}(c)},
[X^{10}]
\right\rangle
+ 
\frac{1}{2^7\cdot 31} {\rm sign} (B_c)\;.
\label{s sp}
\)
The $s$-invariant can be related to the eta-invariants in the adiabatic limit
as follows \cite{DZ}.
Let $\pi: E \to X^{10}$ be an oriented 2-dimensional 
real vector bundle over $X^{10}$ and $g_E$ a fiber metric 
on $E$ with a compatible connection $\nabla^E$.
Let $Y^{11}$ be the unit sphere bundle of $E$ with the induced metric
$g_Y$, so that $Y^{11}$ is a circle bundle over $X^{10}$ with an 
induced Spin structure $\xi$. 
For $\epsilon>0$ consider the metric 
$
g_\epsilon= g_\epsilon^Y= g_E \oplus \pi^* (\frac{1}{\epsilon} g_X)\;.
$
Taking the adiabatic limit, $\epsilon \to 0$, 
gives
\bea
\lim_{\epsilon \to 0} \frac{1}{2}\eta (D(Y^{11}, g_\epsilon^Y))
&=&-
\left\langle 
\widehat{A}(TX^{10}) \left( \frac{1}{e} - \frac{1}{2{\rm tanh} (e/2)}\right)~,~
[X^{10}] \right\rangle\;,
\\
\lim_{\epsilon \to 0} \eta (\cS(Y^{11}, g_\epsilon^Y))
&=&
-\left\langle 
L(TX^{10}) \left(  \frac{1}{{\rm tanh} (e)}
-\frac{1}{e} 
\right)~,~
[X^{10}] \right\rangle- {\rm sign}(B_e)\;.
\eea
Combining the two gives the expression \eqref{s sp} of the $s$-invariant. 
Trading the L-genus with the Rarita-Schwinger index gives back the expressions
derived in \cite{MS} \cite{S-gerbe} \cite{DMW-Spinc}, where the dimensional reduction 
to type IIA string theory is first interpreted via the adiabatic limit. 


\subsection{Disk bundles and the secondary correction term}
\label{sec dis}
We consider the general case when the twelve-manifold no longer
has a product metric near the boundary, which is a departure from the set-up of
APS \cite{APS1}.
Assume then that $(Y^{11}, g_Y)$ bounds a (general) twelve-dimensional Riemannian manifold
$(Z^{12}, g_Z)$. Let $N^{12}=Y^{11} \times [0, 1]$ with product metric $g_0$, 
and extend $g_Z$ smoothly to 
a metric $g_1$ on $Z^{12} \cup N^{12}$ in such a way that $g_1$ is a product 
metric near $Y^{11} \times \{1\}$. Then, from \cite{G},
 the signature of $Z^{12}$ is given by 
\(
{\rm sign} (Z^{12})= \int_{Z^{12}} L_{12} (p_i(g_Z)) + \int_{Y^{11}} TL_{12}(g_0, g_1)
- \eta (Y^{11})\;,
\)
where the boundary correction term $TL_k$ is the secondary characteristic 
$L_k$-class
corresponding to the 
Levi-Civita connections of $g_0$ and $g_1$.
Let $h$ be a fiber metric on the 
line bundle $\cL$ and $\nabla^\cL$
connection on $\cL$ compatible with $h$, so that 
$g_\cL= h + \pi^*(g)$ is an induced natural Riemannian metric
on the total space $\cL$.  Let $S_r(\cL)$ be the circle bundle of 
radius $r$ corresponding to the bundle $\cL$. 
Consider two concentric disk bundles 
$\mathbb{D}_\rho(\cL)$ and 
$\mathbb{D}_r(\cL)$ where $\rho <r$.
Let $g_0(\rho, r)$ be the product metric on the annulus 
$N^{12}=S_r(\cL) \times [\rho, r]$.  Extend the metric $g_\cL$ 
on $\mathbb{D}_\rho(\cL)$ 
to a metric $g_1(\rho, r)$ on $\mathbb{D}_r(\cL)$ in such a way that
$g_1(\rho, r)=g_0(\rho, r)$ near the boundary $S_r(\cL)$.
Let $\nabla^r$ be the Levi-Civita connection for the product metric
$g_0$ on $\mathbb{D}_r(\cL)-\{0\}$ and let $\alpha(r)$ be the corresponding 
connection form.
Let $\nabla^\rho$ be the Levi-Civita connection of $g_\cL|_{\mathbb{D}_\rho (\cL)}$
and $\beta(\rho)$ the corresponding connection form.
Let $\theta=\beta(r)-\alpha(\rho)$ and consider 
$\Omega_t=\Omega_t(\rho, r)$, the curvature of the connection 
$(1-t)\nabla^r + t \nabla^\rho$.
The secondary characteristic $L_{12}$-class is defined in \cite{Ti} by 
\(
TL_{12}(g_0, g_1):= 6\int_0^1 L_{12}(\theta, \Omega_t, \cdots, \Omega_t) dt\;.
\)
Then the signature of the disk bundle is
\(
{\rm sign}(\mathbb{D}_r(\cL))=\int_{\mathbb{D}_r(\cL)}L_{12} (g_1) 
+ \int_{S_r(\cL)} TL_{12}(g_0, g_1)
- \eta (S_r(\cL))\;,
\)
The first term on the right hand side goes to zero as $\rho \to 0$, so that as in 
\cite{Ti} \cite{Ko}
\(
{\rm sign}(\mathbb{D}_r(\cL))=\int_{S(\cL)} \lim_{\rho \to 0} TL_{12}(g_0, g_1) -
\eta(S_r(\cL))\;.
\)
Let $V$ be a real vector bundle of rank 2 over a compact oriented Riemannian
ten-manifold.
Then, using \cite{Ti}, the signature of the disk bundle $\mathbb{D}(V)$   
is
\(
{\rm sign}(\mathbb{D}(V))=\int_{X^{10}} L_{12}(V, X^{10}) -
\lim_{r\to 0} \eta (S_r(V))\;,
\label{th 333}
\)
where $\eta (S_r(V))$ is the eta-invariant of the circle bundle of radius
$r$ and $L_{12}(V, X^{10})$ is the characteristic polynomial of degree ten which is 
expressed explicitly in terms of the coefficients of the Hirzebruch $L_{12}$ 
polynomial, the Euler class $e(V)$ and the Pontrjagin classes $p_i(X^{10})$
\(
L_{12}(V, X^{10})= \frac{1}{3^3\cdot 5 \cdot 7}
\left[  8 e(V)^5 - 14 e(V)^3  p_1(X^{10}) + 49 e(V)  p_2(X^{10})
 -7 e(V)  p_1(X^{10})^2
 \right]\;.
 \label{corr}
\)
Expression \eqref{th 333} shows that $\eta (S_r(V))$ is in general not topological,
and the limiting eta-invariant needs to be included in the expression of the phase when 
considering disk bundles. The geometric correction \eqref{corr} 
occur because we are considering nontrivial circle bundles. Otherwise, when 
$e(V)=0$ we have $L_{12}(V, X^{10})=0$. Also, the expression 
\eqref{corr} would simplify depending on the values of $p_1(X^{10})$ and 
$p_2(X^{10})$, which unlike the ones for $Y^{11}$, we are not assuming to vanish. 
In the case of disk bundles, expression \eqref{corr} is the result in ten dimensions 
of the corresponding expression for the L-genus in twelve dimensions, used  
in our main discussions in section \ref{sec inv}.

\paragraph{Example: Hopf bundle over $\C P^5$.}
Consider type IIA string theory on the complex projective space $\C P^5$. 
For the canonical line bundle $\gamma$ over $\CP^5$, the characteristic 
polynomial is, from \cite{Ti},
\(
L_{12}(\gamma, \C P^5)= 
 \frac{1}{3^3\cdot 5 \cdot 7}
\left[  8 c_1(\gamma)^5 + 14 c_1(\gamma)^3 c_2(\C P^5) 
+ 49 c_1(\gamma) c_4(\C P^5)
 -7 c_1(\gamma) c_2(\C P^5)^2
 \right]\;.
\)
Integrating gives $\int_{\C P^5} L_{12}(\gamma, \C P^5)=-\frac{2^4 \cdot 53}{3^3 \cdot 5 \cdot 7}$. Now with 
${\rm sign}(\mathbb{D} (\gamma))= {\rm sign}(\C P^5)=0$, expression 
\eqref{th 333} gives the value for the limiting eta-invariant
 $\lim_{r \to 0} \eta (S_r(\gamma))=-
\frac{2^4 \cdot 53}{3^3 \cdot 5 \cdot 7}$.

\vspace{3mm}
We hope to make further use of the appearance of the signature and geometric invariants
in M-theory in the near future.


\end{document}